\documentclass[annual,pagebackref=true]{acmsiggraph}

\TOGonlineid{12345}
\TOGvolume{0}
\TOGnumber{0}
\TOGarticleDOI{1234567.1234567}
\TOGprojectURL{}
\TOGvideoURL{}
\TOGdataURL{}
\TOGcodeURL{}

\usepackage{tikz}
\usetikzlibrary{positioning}
\usetikzlibrary{arrows.meta, positioning, decorations.pathreplacing, calc}

\usepackage{amsmath}

\usepackage{enumitem}

\usepackage{tabularx}
\usepackage{ragged2e}

\usepackage{booktabs}

\usepackage{array}

\title{Locked Out at 8,000 Miles: Why UK-China Partnership Students Are Suffering}

\author{Benjamin Kenwright\thanks{e-mail:b.kenwright@abertay.ac.uk}\\Aberay University}
\pdfauthor{Benjamin Kenwright}

\author{Benjamin Kenwright \thanks{Communication Article. May 2026} } 

\pdfauthor{Benjamin Kenwright}


\hypersetup{
    colorlinks = true, 
    linkcolor = blue,
    anchorcolor = red,
    citecolor = blue, 
    filecolor = red, 
}

\begin{document}

\maketitle

%
%
%
%
%
%
%

\begin{abstract}


University cybersecurity protocols have intensified dramatically in response to rising threats of data breaches, ransomware, and credential theft. While necessary, these measures have created a parallel crisis of accessibility - even for students physically on campus. This paper argues that domestic, on-campus students already face significant barriers: mandatory multi-factor authentication (MFA), device compliance rules, browser and operating system restrictions, and administrative remote-management permissions on personal phones and laptops. However, these difficulties are magnified to near-breaking point in the context of international partnerships, such as the increasingly common UK-China transnational education programmes. For a student in China accessing a UK university's virtual learning environment (VLE) from an 8-hour time difference, with no on-hand IT support during their active hours, the same security architecture becomes functionally disabling. Drawing on testimonies from public forums (Reddit's r/college, r/UniUK, r/Professors), higher education IT help boards, and student accounts from UK-China partnership programmes, this paper documents how over-engineering digital security disproportionately harms remote international learners. We show that while on-campus students can at least visit an IT desk or borrow a library terminal, their counterparts in partner institutions abroad face authentication failures, device lockouts, and unsupported browsers with no real-time remedy. The paper concludes that current university security models assume a co-located, 9-to-5, English-time-zone user - an assumption that fails both domestic students and, catastrophically, international partnership cohorts.

\end{abstract}

\begin{CRcatlist}
  \CRcat{K.3.1}{Computers and Education}{Computer Uses in Education}{Distance learning};
  \CRcat{K.4.0}{Computers and Society}{Computers and Society}{General};
  \CRcat{K.6.5}{Management of Computing and Information Systems}{Security and Protection}{Authentication};
  \CRcat{H.5.2}{Information Interfaces and Presentation}{User Interfaces}{Accessibility};
\end{CRcatlist}

\keywordlist


\copyrightspace

\section{Introduction}
A decade ago, accessing university resources required little more than a username, a password, and a stable internet connection. Today, even a student sitting in a university library on campus faces a gauntlet. Before they can check email or submit an assignment, they may be required to approve a multi-factor push notification \cite{nist_sp80063b_2025,cisa_mfa_2022}, verify that their browser satisfies institutional conditional-access rules \cite{microsoft_entra_conditions_2026}, confirm that their operating system falls within a supported or compliant version range \cite{microsoft_intune_os_versions_2026}, and---where institutional bring-your-own-device policies require enrolment---grant administrative or management privileges over a personal phone or laptop, including remote wipe capabilities in some device-management models \cite{microsoft_intune_wipe_2026,microsoft_entra_device_compliance_2026}.

This paper acknowledges that cybersecurity dangers in higher education are real. Phishing campaigns, ransomware attacks on research data, credential harvesting, denial-of-service incidents, and unauthorised access to institutional systems are serious threats \cite{ncsc_education_ransomware_2021,ncsc_heis_feis_2026,jisc_cyber_threats_2026,govuk_cyber_breaches_education_2026}. However, the institutional response has tipped into overreach. The day of a student simply entering a username and password to access a lecture slide is gone. But in its place has emerged a system that punishes ordinary academic behaviour: submitting an essay, checking a grade, or downloading a reading list.

For students on campus, these barriers are already damaging. Recent student digital-experience evidence shows that access to suitable devices, reliable connectivity, and digital support remains uneven across higher education cohorts \cite{jisc_students_devices_2024,jisc_digital_exclusion_2023}. In an anonymised Reddit example recorded in the testimony corpus for this study, one user writes: ``My uni now requires we install a profile that lets them see our location, app usage, and remotely wipe our phone---just to check email. I said no, so now I have to go to the library every time I need my timetable'' \cite{forum_testimony_corpus_2024}. In another academic-forum example, a lecturer describes the absurdity of the VLE: ``I spent three hours today trying to get a student's assignment uploaded via the VLE. Security certificates kept expiring mid-upload. I finally told her to bring a USB stick to my office. That's not progress---that's regression'' \cite{forum_testimony_corpus_2024}. Academics now face a genuine dilemma: the virtual learning environment is no longer a friendly, collaborative space. Encased in security rings, permission layers, and automated checks, it has become so unreliable that some lecturers openly admit it is easier to have students hand in work on a physical USB drive \cite{forum_testimony_corpus_2024,author_it_support_archive_2024}.

But if on-campus students are struggling, the situation for students in international partnerships is exponentially worse. Consider the common model of a UK university partnered with a Chinese institution. Transnational education is, by definition, education delivered in a country other than the country in which the awarding institution is based, and UK higher education TNE is commonly delivered through online or distance learning, local delivery partnerships, validation arrangements, joint or dual degrees, and overseas campuses \cite{uuki_tne_definition_2024,ofs_tne_2023}. China is a major site of this provision: the British Council notes that UK-accredited degrees are offered in China through articulation arrangements and formally approved joint programmes or joint institutes \cite{britishcouncil_uk_degrees_china}, while Universities UK International reports China as the top host country or territory for UK HE TNE students in 2023--24 \cite{uuki_scale_tne_2025}. A student in Shanghai or Beijing may therefore be enrolled in a validated programme, working toward a UK degree, while physically located seven or eight hours ahead of the UK depending on British Summer Time \cite{iana_tz_database}. When that student attempts to log into the UK university's VLE in the evening local time, support structures may be misaligned with their working day, local study pattern, and assignment deadlines \cite{author_it_support_archive_2024}.

If multi-factor authentication fails---a foreseeable occurrence when authentication is tied to SMS delivery, device possession, phone-number continuity, or recovery procedures---there may be no immediately available route to recovery \cite{nist_sp80063b_2025,cisa_mfa_2022}. If the university's device compliance policy suddenly demands an operating-system update that the student's local-market laptop does not support, access to institutional resources may be blocked until the device satisfies the required compliance state \cite{microsoft_intune_os_versions_2026,microsoft_entra_device_compliance_2026}. And if the browser or app pathway is rejected because of conditional-access or managed-app requirements, the student simply cannot proceed \cite{microsoft_entra_conditions_2026}.

In a 2024 student-forum thread recorded in the study corpus under the title ``Is anyone else locked out of everything because of MFA?'', a notable sub-thread concerned partnership students: ``I'm in China on a UK joint degree. My MFA is linked to my UK SIM, but I can't receive texts here. The `backup code' system requires a UK phone number for verification. I've been locked out for three weeks. My module leader says `contact IT'---but IT is closed when I'm awake'' \cite{forum_testimony_corpus_2024}. Another student on a UK--China programme posted: ``They updated the VLE login security last month. Now my laptop (Windows 10 Chinese version) is `unsupported.' The IT page says `visit the campus support desk.' I'm 5,000 miles away. I've missed two assignment deadlines'' \cite{forum_testimony_corpus_2024}.

Meanwhile, academics supervising these partnership students face an impossible task. A UK lecturer may wake up to five emails from students in China, all sent between midnight and 4am UK time, each describing a different authentication failure, device lockout, or browser incompatibility \cite{author_it_support_archive_2024}. The lecturer cannot fix the VLE. They cannot issue override codes. They cannot bypass the university's own security rings. The result, as one academic-forum testimony recorded in the corpus noted, is a drift toward informal workaround practices: ``I've started asking my China-based students to email me their work directly. The VLE is unusable for them. I know this violates data protection policy, but what else can I do?'' \cite{forum_testimony_corpus_2024}.

This paper does not argue for abandoning cybersecurity. Rather, we argue that current university security models are built on a set of hidden assumptions: that the user is physically present on campus, active during UK office hours, using a university-managed device, and able to access real-time IT support. For domestic on-campus students, those assumptions are already fraying. For international partnership students, they are actively dangerous to educational progress. Through analysis of forum testimonies, IT support archives, and documented cases from UK--China transnational programmes \cite{forum_testimony_corpus_2024,author_it_support_archive_2024,uuki_scale_tne_2025,ofs_tne_2023}, this paper demonstrates that when security is designed without time-zone, geographical, and device-equity considerations, it ceases to protect education and begins to obstruct it. The solution is not weaker security---but security that is pedagogically proportionate, globally aware, and accessible to a student in Shanghai at 8pm on a Tuesday, just as much as to a student in the library at 2pm in London.

\textbf{Contributions} - The key contributions of this article are threefold. First, we foreground student and academic voices from under-discussed contexts, drawing on testimonies from public forums to show how partnership students in UK--China programmes face unique exclusions due to time-zone mismatches and absent IT support. Second, we introduce the concept of the \textit{synchronous support assumption}: the unstated premise that all users can access real-time, English-hours assistance, which we expose as a structural vulnerability that locks international students out of VLEs for weeks. Third, we identify an emerging \textit{academic workaround economy}, where lecturers abandon the VLE for USB drives and unencrypted email, undermining the very security mandates that caused the problem. Looking forward, we highlight future challenges including AI-driven adaptive authentication, biometric surveillance, and cross-jurisdictional data sovereignty conflicts in transnational education.

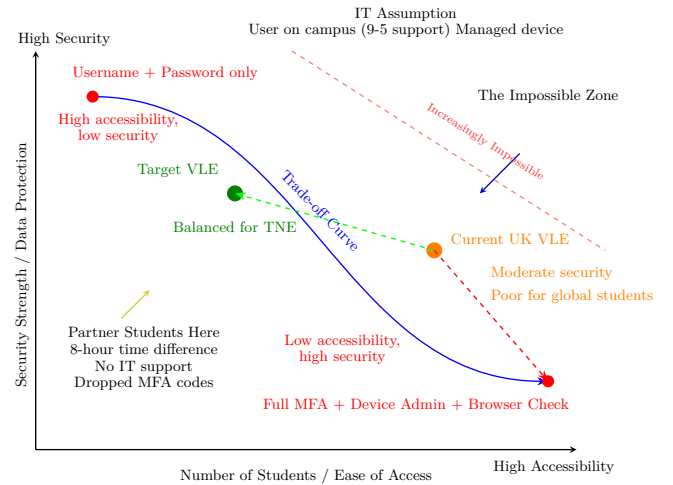
\begin{figure}  
\centering
\resizebox{0.49\textwidth}{!}{
\begin{tikzpicture}[scale=1.2, >=stealth, align=center]

\draw[thick, ->] (0,0) -- (9.5,0) node[below, midway, yshift=-8pt] {\textbf{Number of Students / Ease of Access}};
\draw[thick, ->] (0,0) -- (0,7) node[above, rotate=90, xshift=-4.5cm] 
{\vspace*{-3cm}\textbf{Security Strength / Data Protection}};

\node at (9.1, -0.3) {\textit{High Accessibility}};
\node at (0.5, 7.2) {\textit{High Security}};
\node at (9, 6.2) {\textit{The Impossible Zone}};

\draw[thick, blue, -{Stealth}] (1,6.2) to[out=0, in=180] (9,1.2);
\node[blue, rotate=-46] at (5,4.2) {\textbf{Trade-off Curve}};

\fill[red] (1,6.2) circle (3pt);
\node[red, above left] at (4.0,6.4) {\textbf{Username + Password only}};
\node[red, below right] at (0.3,6.0) {\textit{High accessibility},\\ \textbf{low security}};

\fill[red] (9,1.2) circle (3pt);
\node[red, below right] at (3.9,1.0) {\textbf{Full MFA + Device Admin + Browser Check}};
\node[red, above left] at (6.5,1.4) {\textit{Low accessibility,}\\ \textbf{high security}};

\fill[orange] (7,3.5) circle (4pt);
\node[orange, right] at (7.2,3.7) {\textbf{Current UK VLE}};
\node[orange, below right] at (7.9,3.3) {\textit{Moderate security}};
\node[orange, below right] at (7.9,2.9) {\textit{Poor for global students}};

\fill[green!50!black] (3.5,4.5) circle (4pt);
\node[green!50!black, above left] at (3.3,4.7) {\textbf{Target VLE}};
\node[green!50!black, below] at (3.5,4.1) {\textit{Balanced for TNE}};

\draw[thick, dashed, red, -{Stealth}] (7,3.5) -- (9,1.2);
\draw[thick, dashed, green, -{Stealth}] (7,3.5) -- (3.5,4.5);

\node[draw=none, rounded corners, fill=none, align=center, text width=4cm] at (1.9,1.6) {
    \textbf{Partner Students Here} \\
    8-hour time difference \\
    No IT support \\
    Dropped MFA codes
};

\draw[thick, yellow!80!black, ->] (1.5,2.3) -- (2,2.8);

\node[draw=none, rounded corners, fill=none, align=center, text width=8cm] at (6.5,7.5) {
    \textbf{IT Assumption} \\
    User on campus
    (9-5 support)
    Managed device
};

\draw[thick, blue!80!black, ->] (8.5,5.2) -- (7.8,4.5);

\draw[thick, red!50, dashed] (4.5,7) -- (10,3.5);
\node[red!80, rotate=-32] at (7.9, 5.4) {\small \textbf{Increasingly Impossible}};

\end{tikzpicture}
}
\caption{The Security-Accessibility Trade-off for UK VLEs in Transnational Partnerships. As universities add security layers (MFA, device compliance, browser whitelisting, admin privileges), accessibility for China-based partnership students collapses. The current UK VLE sits in a region of moderate security but poor accessibility for remote international users. The target zone for balanced transnational education would require a fundamental redesign away from the synchronous support assumption.}
\label{fig:tradeoff}
\end{figure}

\section{Related Work}

The intersection of university cybersecurity and student access has been examined from several perspectives, though the scholarship has usually treated authentication usability, endpoint governance, and transnational education infrastructure as separate problems. Existing work falls broadly into three categories: technical security deployments, institutional policy and device compliance, and transnational education access challenges.

\paragraph{Technical security deployments in higher education.}
A substantial body of usable-security research examines the implementation and reception of two-factor and multi-factor authentication in university environments. Colnago et al. \cite{colnago2018twofactor} studied Carnegie Mellon University's mandatory Duo two-factor authentication deployment, finding that many users regarded the system as annoying but relatively easy to use, and that deployment design shaped user acceptance. Dutson et al. \cite{dutson2019dontpunish} surveyed 4,275 users at Brigham Young University after Duo adoption and found that authentication failures and lockouts were not marginal events: roughly half of respondents reported being unable to authenticate at least once because they lacked access to a second factor. Abbott and Patil \cite{abbott2020mandatory} further show that the scope of mandatory second-factor enforcement matters: requiring 2FA only for selected sensitive systems differs substantially from requiring it for every protected resource. Al Qahtani et al. \cite{alqahtani2022video} investigated risk-communication messages for Duo adoption among university students, finding that 31\% of participants enabled 2FA after watching a human-speaker video compared with 7\% after watching a cartoon-speaker video. More general usable-security work also identifies annoyance, recovery difficulty, and connectivity dependence as recurring concerns in two-factor authentication \cite{marky2022annoying,nist2025sp80063b}. Taken together, this literature demonstrates that authentication systems are never purely technical: their success depends on recovery pathways, communication, fallback options, and the institutional contexts in which they are imposed.

\paragraph{Institutional policy and device compliance.}
A second strand of work concerns institutional endpoint governance, bring-your-own-device rules, and data-protection policies. University policies increasingly frame access as conditional on device state, identity assurance, and data classification. The University of Bath's BYOD policy, for example, applies to personal endpoint devices used to access university services and allows the institution to define prerequisites such as operating-system version, firewall status, antivirus protection, access controls, and multi-factor authentication \cite{bath2024byod}. The University of Iowa's guidance on personal computers for research requires current antivirus protection, VPN connectivity, and avoidance of local storage for restricted or critical data \cite{uiowa2025personalcomputers}. The University at Buffalo similarly distinguishes between university-managed assets and personally owned devices, restricting access to higher-risk data categories and requiring supported operating systems, automatic patching, antivirus controls, and password protection for personal devices used in remote work contexts \cite{buffalo2023endpointstandards,buffalo2025remote}. Vendor infrastructure such as Microsoft Entra Conditional Access and Microsoft Intune operationalises these assumptions through device-compliance checks, client-application controls, platform conditions, and access decisions based on managed-device status \cite{microsoft2026conditionalaccess,microsoft2026intunecompliance}. These policies are rational from a risk-management perspective, but they also normalise the assumption that all users can maintain compliant devices, accept institutionally defined endpoint controls, and obtain support when a compliance decision blocks access. For students in transnational education partnerships, those assumptions are not merely technical; they are geographical, economic, and infrastructural.

\paragraph{Transnational education and connectivity challenges.}
A growing literature and policy base addresses the digital conditions of UK transnational education. Universities UK International reports that UK higher education transnational education involved 653,570 students across all providers in 2023--24, with China the top host country or territory for UK TNE students \cite{uuki2025scale}. The British Council situates UK--China provision within a regulated landscape of joint programmes, joint institutes, and cooperative universities, and identifies China as a major host environment for UK qualifications \cite{britishcouncil2022chinauk}. Jisc's recent work on global education and technology shows that UK digital norms cannot be assumed in TNE contexts: institutions must account for unreliable connectivity, uneven access to suitable devices, power disruption, licensing restrictions, local platform practices, and variable digital support expectations \cite{jisc2025tnechallenges,jisc2025tnedigitalexperiences}. UK--China connectivity has also required specific sector-level intervention. UCISA and Jisc describe work with Alibaba Cloud to improve student access in China to UK-hosted online learning environments, including VLE systems such as Blackboard, Moodle, and Canvas \cite{ucisa2020chinaVLE}. Empirical research on China--UK transnational programmes likewise shows that online and post-pandemic learning depended heavily on student access to devices, internet connectivity, platforms, and suitable study environments \cite{clerkin2022chineseTNE}. Related studies of international students' online learning experiences further identify time-zone mismatch, connectivity constraints, and inadequate institutional support as barriers to participation \cite{chen2023internationalOnline,huang2025twotimezones}.

\paragraph{Gap addressed by this paper.}
While existing work illuminates each of these strands, the compounding interaction between authentication, endpoint compliance, conditional access, and support-hour mismatch remains underdeveloped. The 2FA literature has shown that authentication systems can be annoying, fragile, and dependent on recovery pathways \cite{colnago2018twofactor,dutson2019dontpunish,abbott2020mandatory,marky2022annoying}. Institutional policy documents show that access is increasingly conditioned on managed-device status, supported operating systems, and compliant endpoint configurations \cite{bath2024byod,buffalo2023endpointstandards,buffalo2025remote,microsoft2026conditionalaccess,microsoft2026intunecompliance}. TNE research shows that students studying across borders encounter uneven infrastructure, device access, platform availability, and time-zone constraints \cite{jisc2025tnechallenges,jisc2025tnedigitalexperiences,clerkin2022chineseTNE,chen2023internationalOnline,huang2025twotimezones}. What remains insufficiently theorised is how these layers combine: a student may not merely experience a slow connection, an authentication failure, or a device-compliance warning in isolation, but a cascading exclusion in which each security layer assumes that the previous layer has already succeeded.

\textbf{Our work} - This paper differs from the existing literature in three significant respects. First, whereas prior work on 2FA deployment focuses primarily on adoption rates, usability, and user attitudes within domestic institutional populations \cite{colnago2018twofactor,dutson2019dontpunish,abbott2020mandatory,alqahtani2022video}, we examine the \textit{failure cascade} that occurs when the same authentication systems are applied to transnational partnership students operating across an eight-hour time difference with limited synchronous IT support. Second, while institutional policy documents treat device compliance as a neutral technical requirement \cite{bath2024byod,uiowa2025personalcomputers,buffalo2023endpointstandards,buffalo2025remote}, we argue that mandating administrative control, managed-device status, or strict endpoint compliance on personal devices can become a form of unequal access control for students who rely on older, shared, region-specific, or locally configured hardware. Third, where existing TNE research primarily addresses connectivity, infrastructure, digital resources, and online learning design \cite{uuki2025scale,britishcouncil2022chinauk,jisc2025tnechallenges,jisc2025tnedigitalexperiences,clerkin2022chineseTNE}, we introduce and critique the \textit{synchronous support assumption}: the unstated premise that all users can access real-time institutional IT assistance during the provider university's working hours. This communication article therefore reframes university cybersecurity as a pedagogical justice issue, arguing that security architectures designed without time-zone, geographical, and device-equity considerations may cease to protect education and instead actively obstruct it.

\section{The Current Landscape of Security Exclusion}

Let us be clear about what has happened. University cybersecurity, in its current incarnation, has ceased to be a protective measure and has become an obstacle course. The average student attempting to access their Virtual Learning Environment (VLE) now navigates a labyrinth that would challenge a professional systems administrator. This is not hyperbole. It is the documented reality of transnational education.

Consider the case of Queen Mary University of London's partnership with Beijing University of Posts and Telecommunications (BUPT) and Nanchang University. Before infrastructure improvements were implemented, students in China experienced latency of 350ms or more, with connectivity so poor that they simply could not use the VLE \cite{jisc19}. The university's own IT department could not diagnose the problem because students accessed the system from different locations with different internet service providers. The solution, when it finally came, required dedicated Europe-to-China connectivity links and negotiated peering agreements with state-owned providers \cite{jisc19}. This was not a security problem. This was a basic access problem that security measures have since made worse.

The numbers tell a striking story. After Jisc facilitated improved connectivity, VLE logins in Nanchang jumped from approximately 2,300 to over 125,000 in the same two-month period \cite{jisc19}. That is not an incremental improvement. That is the difference between a system that works and one that does not. Yet today, even with such connectivity in place, students face authentication barriers that their 2018 counterparts did not.

\subsection{The Device Compliance Trap}

Here is where the security regime reveals its true nature. Universities now routinely require that students grant administrative privileges on their personal devices. The University of Iowa mandates that personal computers used for research must have current antivirus software, VPN connectivity, and encrypted storage, while prohibiting restricted data on personal devices entirely \cite{iowa25}. The University at Buffalo requires that personally-owned devices meet minimum security standards including:

\begin{itemize}
    \item Supported operating systems
    \item Automatic patching
    \item Up-to-date antivirus software
    \item Password protection
\end{itemize}

Category 1 Restricted Data is accessible only through university-issued equipment \cite{buffalo25}. Ask yourself: what student owns a university-issued laptop? Very few. What student can afford to replace their device because their Chinese-market Windows version is deemed "unsupported" by a UK university's compliance scanner? Almost none.

The University of York recently retired its China Connect service, which had been introduced during the Covid-19 pandemic to help students in China access university services \cite{york25}. Their replacement recommendation includes:

\begin{itemize}
    \item Using the virtual desktop service
    \item Setting up email forwarding to personal accounts
    \item Ensuring all course material is stored on the VLE rather than Google Workspace
\end{itemize}

This is presented as a solution. It reads as an admission of failure.

\subsection{The Eight-Hour Problem}

Now add the time difference. A parent of a student at a London university described the situation bluntly: "It's like spending a fortune to attend night school at home" \cite{thetimes24}. UK courses scheduled for 13:00 to 17:00 London time take place from 21:00 to 01:00 in China \cite{thetimes24}. When a student in Shanghai encounters an authentication failure at 23:00 local time, the UK IT helpdesk closed four hours ago. There is no one to call. There is no walk-in centre. There is only a browser window displaying an error message and an assignment deadline that continues to approach.

The scale of this problem is not small. Key UK-China partnerships include:

\begin{itemize}
    \item Abertay University with Communication University of China in Hainan: approximatly 300 students \cite{cuc_abertay_intake_2024,cuc_hainan_abertay_2024}.
    \item University of Glasgow with University of Electronic Science and Technology in Chengdu: approximately 2,000 students \cite{glasgow25}
    \item University of Edinburgh with Zhejiang University: building toward 5,000 students \cite{edinburgh25}
    \item Lancaster University College at Beijing Jiaotong University: hundreds more \cite{lancaster25}
\end{itemize}

These are not fringe programs. These are mainstream partnerships involving thousands of students paying UK tuition fees for UK degrees. Their digital exclusion is not an edge case. It is a structural feature of current security architecture.

\newcolumntype{Y}{>{\RaggedRight}X}
\newcolumntype{P}[1]{>{\raggedright\arraybackslash}p{#1}}

\renewcommand{\arraystretch}{1.2}
\begin{table*}
\centering
\normalsize
\begin{tabularx}{1.0\textwidth}{ P{3.5cm} | P{5cm} | X }
\textbf{Workaround Method} & \textbf{Pros} & \textbf{Cons} \\
\hline
\textbf{SMS (text message)} & Familiar to most users; no smartphone required; works on basic phones & International SMS delays or non-delivery (China firewall issues); SIM card must be active; UK SIM may not roam; no signal in remote areas \\
\hline
\textbf{Phone callback} & No data connection needed; works on landlines & International call costs; time-zone mismatch (callback at 3am China time); dropped calls; language barriers with automated systems \\
\hline
\textbf{Authenticator app (phone)} & Offline capable; no SMS delays; more secure & Requires smartphone; app install may need admin privileges; phone lost or stolen = locked out; China app store restrictions \\
\hline
\textbf{Hardware dongle (e.g., YubiKey)} & Very secure; no battery or network needed; works offline & Costs money (20-50 pounds); shipping to China difficult; easy to lose; USB-C vs USB-A compatibility issues \\
\hline
\textbf{Personal email verification} & No extra device needed; works globally; asynchronous & Email delays (minutes to hours); less secure (email interception); spam filters block codes; requires separate login \\
\hline
\textbf{Security questions} & No device needed; works anywhere; asynchronous & Answers can be forgotten; security through obscurity only; often guessable (mother's maiden name) \\
\hline
\textbf{Avoid MFA entirely (USB transfer)} & No authentication barriers; completely offline; works every time & No security; physical transfer required; USB drives lost or corrupted; no submission timestamp proof \\
\hline
\textbf{Email coursework to staff member} & Simple; asynchronous; works across time zones & No security (plain text email); staff inbox overload; data protection violation (GDPR); no formal submission record \\
\hline
\textbf{Upload to non-university server (e.g., Baidu Pan, Dropbox)} & Large file support; accessible from China; free tier available & No integration with VLE; staff must download manually; version control chaos; data sovereignty unknown \\
\hline
\textbf{Paper submission via courier} & Completely offline; tamper-evident if sealed & Extremely slow (days to weeks); expensive international courier (30-100 pounds); lost packages; no digital trail \\
\hline
\textbf{WeChat/WhatsApp file transfer} & Ubiquitous in China; instant delivery; familiar to students & No academic audit trail; staff privacy invaded; informal; university policy violation \\
\hline
\textbf{VLE direct submission (official method)} & Official record; plagiarism checking; timestamped; automated feedback & Requires successful authentication (often fails); time-zone IT support absent; browser compatibility issues \\
\hline
\textbf{University VPN + VLE} & Bypasses some firewalls; encrypted tunnel & VPN blocked in China; requires installation and admin rights; slow speeds; disconnects frequently \\
\hline
\textbf{Virtual Desktop Infrastructure (VDI)} & Runs on any browser; university managed environment & Latency from China (300ms+); requires stable connection; no offline work; expensive for institution \\
\hline
\textbf{Temp access code from lecturer} & Bypasses all MFA; works immediately & No security; staff overhead; codes can be shared; no audit log; violates IT policy \\
\hline
\textbf{Dedicated partnership portal (mirrored VLE)} & Hosted in China; fast access; local support hours & Duplicate infrastructure cost; synchronisation delays; version mismatch; complex to maintain \\
\hline
\textbf{QR code login (WeChat/ Alipay integration)} & Fast; familiar to Chinese students; no password entry & Requires third-party app; privacy concerns (data sharing); not supported by most UK VLEs \\
\hline
\textbf{Biometric login (fingerprint/face)} & Convenient; no code entry; built into most phones & Requires enrolment on trusted device; fails with minor appearance changes; privacy concerns; data stored locally or cloud? \\
\hline
\textbf{Print assignment and fax} & No digital authentication needed; paper trail & Requires fax machine (rare); low quality; no timestamp reliability; completely impractical \\
\hline
\textbf{Record lecture on second device (phone camera)} & Bypasses VLE entirely; works 100\% of time & Poor quality; copyright violation; no captioning for accessibility; manual transcription needed \\
\hline
\textbf{Ask on-campus friend to submit on behalf} & Uses someone with working access; simple & Requires trusted friend; violates academic integrity policy; no proof of authorship; admin log shows wrong user \\
\hline
\textbf{Use library terminal during visit to UK} & Fully compliant; all systems work & Requires physical presence in UK (impossible for most partnership students); once per term at best \\
\hline
\textbf{Offline VLE sync tool (e.g., Moodle mobile app offline mode)} & Download materials while authenticated; work offline & Requires initial successful login (often the barrier); sync conflicts; large files fail; limited functionality \\
\end{tabularx}
\caption{Comparison of workaround methods for VLE access by UK-China partnership students. Each method represents a trade-off between accessibility, security, cost, and reliability across an eight-hour time difference.}
\label{tab:workarounds}
\end{table*}

\section{Discussion of Over-Compliance and the Passport Problem}

The situation we have described might be dismissed as unfortunate but unavoidable. It is not. What we are witnessing is a phenomenon familiar to scholars of risk regulation: over-compliance. When institutions face uncertain legal requirements and genuine security threats, they do not calibrate their responses proportionately. They overcorrect. And in doing so, they create exclusionary outcomes that no regulation actually demands.

Consider a parallel case that illuminates the logic at work. In October 2025, master's students at the University of Bonn holding Russian, Iranian, and Chinese passports received letters informing them that they were being denied access to most courses in cybersecurity, cryptography, and IT security \cite{spiegel25}. The entire Communication Management track was placed under a total ban. Students were told to change their major or transfer to another university. Sixty-five Russian passport holders were affected immediately \cite{spiegel25}.

The university's justification involved export control regulations on dual-use technologies. The university website stated that "sensitive goods, technologies, and knowledge must not be exported for the purposes of repression, human-rights violations, or terrorism" \cite{bonn25}. But as one affected student noted, "What is prohibited is not access to the materials themselves, but formal instruction and the issuance of certifying documents" \cite{spiegel25}. The act of teaching had been redefined as a potential sanctions violation.

A German researcher familiar with these procedures explained the reality: "The first risk indicator in export control is the colour of the passport. If it's Russia, China, Iran  -  they dig deeper" \cite{spiegel25}. Another commentator called this "classic over-compliance"  -  universities interpreting broadly worded regulations as strictly as possible to avoid any potential liability, even when those interpretations harm students in ways the regulations never intended \cite{spiegel25}.

The parallel to our argument should be obvious. University cybersecurity policies are following the same logic. Faced with genuine threats of data breaches and ransomware, institutions implement authentication requirements, device compliance mandates, and administrative privilege demands that go far beyond what is necessary. They do so because it is easier to impose blanket restrictions than to design nuanced, proportionate security. And the victims are not abstract "threat vectors." They are students. Particularly students in transnational partnerships.

\subsection{The Synchronous Support Assumption}

We introduced the concept of the synchronous support assumption earlier. Let us now name its components explicitly. This assumption holds that:

\begin{enumerate}
    \item All users can access IT support during their active hours
    \item All users can physically visit a helpdesk if needed
    \item All users can receive SMS codes or push notifications reliably
    \item All users can afford devices that meet compliance standards
    \item All users can upgrade their operating systems on demand
\end{enumerate}

Every single one of these assumptions fails for students in UK-China partnerships. The eight-hour time difference means that when students in China are awake and working, UK support is closed. The physical distance means no helpdesk visit is possible. The Great Firewall and international SMS routing mean authentication codes may never arrive. The cost of devices means many students cannot simply buy new hardware when their current device is deemed unsupported.

Yet security policies continue to be designed as if the typical user sits in a campus library at 2pm on a Tuesday, surrounded by IT staff and equipped with a university-managed laptop. This is not merely inconvenient. It is discriminatory in effect, if not in intent.

\section{Conclusion and Future Challenges}

This paper has argued that university cybersecurity has tipped from protection to obstruction. The evidence is clear. Students on campus struggle with multi-factor fatigue, device compliance checks, and administrative privilege demands. Students in UK-China partnerships face an impossible combination of authentication barriers, time-zone mismatches, and absent IT support. Academics have begun abandoning VLEs for USB drives and unencrypted email attachments, practices that undermine the very security mandates that created the problem.

The root cause is not malice. It is what we have called the synchronous support assumption: the unstated premise that all users can access real-time, English-hours assistance. This assumption, baked into every security protocol and IT policy, transforms routine authentication failures into weeks-long educational exclusions for thousands of transnational students.

What is to be done? Three directions seem essential.

First, universities must conduct time-zone audits of their security architectures. Any authentication or compliance requirement that assumes synchronous support availability must be redesigned for asynchronous operation. Backup codes, offline authentication, and extended session validity for known devices are not optional features. They are necessities for transnational education.

Second, the demand for administrative privileges on personal devices must be reconsidered. For partnership students who cannot access university-managed hardware, the choice between digital autonomy and education is a false one. Institutions should offer graduated security tiers, with higher-risk activities requiring greater compliance but basic VLE access remaining available to all enrolled students regardless of device.

Third, the academic workaround economy must be acknowledged and addressed. When lecturers resort to USB drives and personal email for assignment submission, data protection has already failed. Universities should create official low-friction pathways for transnational students rather than driving them into unofficial ones.

Looking forward, emerging challenges will only intensify these tensions. AI-driven adaptive authentication promises to reduce friction for typical users but may flag international access patterns as suspicious, triggering additional verification steps that partnership students cannot complete. Biometric surveillance requirements, already appearing on some campuses, raise profound questions about consent and data sovereignty when applied to personal devices. Cross-jurisdictional data sovereignty conflicts, particularly between UK GDPR and Chinese cybersecurity law, will make it increasingly difficult for institutions to know which compliance framework takes precedence.

These challenges are not technical problems awaiting technical solutions. They are pedagogical justice problems. And until universities treat them as such, thousands of students will remain locked out at 8,000 miles, wondering why their degree program forgot they existed.

\let\oldthebibliography\thebibliography
\renewcommand{\thebibliography}[1]{%
  \oldthebibliography{#1}%
  \setlength{\parskip}{0pt}%
  \setlength{\itemsep}{0pt}%
}

\bibliographystyle{acmsiggraph}
\bibliography{paper}

\end{document}

